\documentstyle[epsf,seceq,mbf,amsfonts,amssymb]{ptptex}

\def\ben{\begin{equation}}
\def\een{\end{equation}}
\def\bena{\begin{eqnarray}}
\def\eena{\end{eqnarray}}

\title{\bf Brane collisions and braneworld cosmology } 

\author{Uchida Gen, Akihiro Ishibashi, and Takahiro Tanaka} 

\inst{Yukawa Institute for Theoretical Physics, 
Kyoto University, Kyoto 606-8502}

\abst{
Recently, we proposed a new braneworld model~\cite{GIT} in which 
the collision of a brane universe and a vacuum bubble coming from 
the extra-dimension is utilized as a trigger of brane big-bang. 
In this article, mainly reviewing this model, 
we briefly summarize cosmological braneworld scenarios 
in which collision of branes plays an important role. 
} 

\begin{document}

\maketitle

\section{Introduction} 

Braneworld scenario~\cite{HW} especially of the Randall-Sundrum 
type~\cite{RS} have recently attracted much attention. 
In particular, cosmological models in this scenario 
have been studied actively~\cite{BWC,CP}. 
It has been shown that braneworld cosmology seems to be consistent with 
the $4$-dimensional conventional cosmology at least 
on scales much lower than that corresponding to the brane tension,   
provided bulk Weyl curvature is sufficiently small. 
However, it is not clear whether braneworld cosmology can 
predict any evidences for the existence of extra-dimension(s) 
which are testable in near future observations. 
Further, it is still not evident whether 
braneworld models actually have a great advantage over the conventional ones. 
Under such a current situation, it is an important direction of research 
to seek for an alternative scenario in which 
the existence of extra-dimension(s) plays an essential role.

As one of such attempts, the present authors recently proposed a new 
cosmological braneworld model~\cite{GIT} in which an inflation occurs 
on a boundary brane driven by small mismatch between the bulk vacuum energy 
and the brane tension, and the nucleation of a true vacuum bubble 
becomes a trigger of the big-bang in the braneworld.   
One of the distinctive features in our model is 
that the bubble nucleation occurs in extra-dimension(s). 
Not only does such a vacuum bubble coming from extra-dimension  
heat up the brane universe through the colliding process, 
but also provide simultaneously an anti-de Sitter bulk of 
the Randall-Sundrum setup, reducing the effective cosmological 
constant on the brane to zero. 

In this article, we shall briefly summarize a colliding brane cosmology. 
We first give a brief review of colliding brane models 
so far proposed in the next section. 
Then, in section~\ref{sect:Bbbbbbb}, we illustrate our brane big-bang 
model proposed in Ref.~\cite{GIT}. In section~\ref{Sect:Discussion}), 
we discuss problems in our model, some of which may be common 
in any types of colliding brane models. 

\section{Colliding brane cosmology} 
\label{Sect:brane-collision} 

Recently there have appeared several interesting works 
in which collision of branes is actively used. Among them, 
the idea called brane inflation, 
in which the interbrane separation plays the role of an inflaton field,
is rather promissing to work~\cite{braneinflation}.  
On the other hand, one of the most 
ambitious proposals is the ekpyrotic 
model~\cite{KOST}, which aims at solving major 
cosmological problems without the use of inflation.   
A severe criticism~\cite{KKL} to it and a number of calculations 
of density fluctuations with the controversial 
claim have also been made~\cite{ControversialKOST}. 

Also in the context of Horava-Witten theory, an intriguing possibility 
that the collision of a visible brane and a bulk moving brane 
generates the baryon asymmetry on a visible 
brane universe has been proposed very recently 
by~Bastero-Gil et.al~\cite{BGCGLP}.

Within the context of the Randall-Sundrum scenario, some models to consider 
bubble nucleation in the bulk have been discussed in several different 
contexts~\cite{bubblenucleation}. 
For example, an idea to realize the Randall-Sundrum setup by a collision 
of bubbles was discussed by Gorsky and Selivanov~\cite{Gorsky}, 
where the bubbles nucleate through the Schwinger process 
in some external field. 

Bucher~\cite{Bucher} proposed an interesting 
model in which anti-de Sitter bubbles appear as a result of a false vacuum 
decay~\cite{CD} and a collision of the two nucleated bubbles 
create a hot big bang universe, giving a possible origin 
of a homogeneous and isotropic bulk and brane geometry.  
Density perturbations in this model have also been calculated~\cite{JB,GT2} 
to show that the scale-invariant spectrum does not easily arise.    
But the result does not immediately exclude the possibility of 
this model because the amplitude due to the effect of the bubble 
wall fluctuations tends to be very tiny. 
Our model we review in the next section has several similarities with 
this colliding bubble model. 

It should be commented that in general relativity, brane (shell) 
collisions have been discussed in a number of literatures~\cite{collision}.  
The formalism for treating collision of gravitating shells 
developed in conventional general relativistic context 
has been extended to more general cases with an eye for applications 
in braneworld cosmology~\cite{LMW}. 
However, concerning perturbations of colliding brane models, 
as far as the present authors know, 
such a formalism taking self-gravity of colliding branes into account 
has not been developed yet.  
Most of perturbation calculations in colliding brane models have been made 
by ignoring self-gravity of colliding branes, or after reducing the system in 
question to effective $4$-dimensional theories with a scalar field 
which mimics fluctuation of a moving brane.

\section{Brane big-bang brought by bulk bubble}  
\label{sect:Bbbbbbb}

Now we shall illustrate our idea~\cite{GIT}.  
In our model, 5-dimensional bulk spacetime is supposed to nucleates 
in a false vacuum phase with a single positive tension brane 
at the fixed point of ${\mathbb Z}_2$-symmetry. 
The false vacuum bulk can be locally Minkowski or de Sitter space. 
The pre-existing brane is in an inflationary phase because of 
the mismatch between the bulk vacuum energy and the brane tension.   
This inflationary phase would last forever if there were 
no mechanism to terminate it. 
However, since the bulk is initially in a false vacuum state, 
a true vacuum anti de Sitter-bubble (AdS-bubble) spontaneously 
nucleates in the bulk as a result of the false vacuum decay 
via quantum tunneling~\cite{CD}.  
This AdS-bubble expands in the false vacuum bulk. 
If the transition occurs with the highest symmetry, 
the nucleated bubble has the common center which respects 
the symmetry of the bulk-brane system. 
However, even if the transition with the highest symmetry is 
the most probable process as discussed in Ref.~\cite{CGM}, 
quantum fluctuations lead to displacement of the position of 
the nucleation from the center of the symmetry. 
Then, because the surface of the Ads-bubble expands just like a 
de Sitter space, the bubble eventually hits the inflationary brane universe. 
The point is that the intersection of the brane and the bubble 
is spacelike. Thus, when the bubble hits the brane, 
the energy of the bubble wall can be converted into radiation on the brane 
unless it dissipates into the bulk. 
Furthermore, 
the effective cosmological constant on the brane is reduced 
with the true vacuum energy chosen to be the negative value which 
balances the tension of the brane. 
As a result, the inflation comes to an end, and the brane can be thermalized 
through this colliding process. 
It is worth noting that the brane is instantaneously heated up at 
the colliding surface beyond the horizon scale of the brane. 
Although such a type of thermalization appears a causality violation 
from the viewpoint of the observers on the brane, 
it is a natural consequence of the bubble nucleation 
in the bulk (outside the brane). We call this collision hypersurface 
a ``~big-bang surface.~''  
If the brane inflation lasts long enough before the collision, 
the big-bang surface can become homogeneous and isotropic. 
Then, in the future of the big-bang surface, the brane evolves as a radiation  
dominated Friedmann-Lemaitre-Robertson-Walker~(FLRW) universe. 

\begin{figure}  
\centerline{\epsfxsize = 9.5  cm \epsfbox{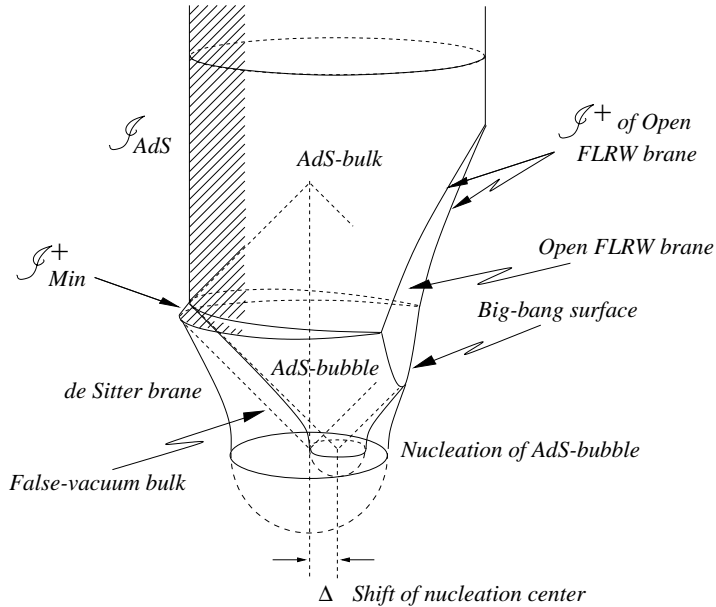}}
\vspace{3mm}
\begin{center}
\begin{minipage}{14cm}
        \caption{\small   
          The conformal diagram shows the brane big-bang scenario. 
          The dotted line hemisphere represents the instanton of the system. 
          A true vacuum AdS-bubble is nucleated and expands in the false 
          vacuum bulk bounded by an inflating de Sitter brane. 
          The expanding AdS-bubble eventually collides with a portion of 
          the de Sitter brane. The case that the separation $\Delta$ of 
          the two centers is spacelike is illustrated.   
          In this case, the intersection, i.e., 
          the big-bang surface, has a hyperbolic geometry ${\mathbb H}^3$ 
          and an open FLRW brane universe is realized after the brane 
          big-bang. This geometry is glued along the boundary surfaces, 
          except ${\cal I}_{AdS}$ and ${\cal I}^\pm_{Min}$, 
          onto a copy of itself with ${\mathbb Z}_2$-symmetry being satisfied. 
                  }        \protect \label{fig:geom}  
\end{minipage}
\end{center}
\end{figure}

After the collision, the bulk around the brane becomes anti-de Sitter 
spacetime and the gravity is effectively localized on the brane by 
the Randall-Sundrum mechanism. 
Since the true vacuum energy is lower than that in the false vacuum, 
this model allows a creation of anti-de Sitter bulk from de Sitter or 
Minkowski-bulk~\cite{CD}.  
The whole story is summarized in Fig.~\ref{fig:geom}. 

In this model the type of the resultant FLRW brane universe 
depends on the location of the bubble nucleation in the bulk; 
It can be open, closed, or flat, if the separation $\Delta$ 
between two centers of the bubble and the brane 
is in spacelike, timelike, or null separation, respectively. 
Furthermore it can be shown that, for the open and closed FLRW brane 
universe, the spatial curvature radius $a_i$ at the moment of 
the brane big-bang is related to the magnitude of $\Delta$ and 
the brane's curvature 
radius $\alpha_B$ as $|\Delta| a_i \approx \alpha_B \ell$ for 
case (a): $\ell^{-2} \gg \alpha_B^{-2}$, and 
$|\Delta| a_i \approx \alpha_B^2$ 
for case (b): $|\ell^{-2}| \lesssim \alpha_B^{-2}$, with $\ell$ 
being the curvature radius of the false vacuum bulk. 
It turns out that in order to solve the flatness problem, 
$\Delta$ must satisfy 
\ben 
  { |\Delta| \over \alpha_B}  
      \lesssim  
  10^{-32} \left( \frac{\sigma_W \ell^5}{\alpha_W} \right)^{1/4}\,,  
\,\,\,\, 
  { |\Delta| \over \alpha_B}  
      \lesssim   
  10^{-32} \left( \sigma_W\alpha_B^4\frac{\alpha_B}{\alpha_W}\right)^{1/4}\,, 
\label{condi:flatness}  
\een  
for case (a) and case (b), respectively. Here $\alpha_W$ and $\sigma_W$ 
denote the bubble wall's radius and tension. 
Thus, for the flatness, the bubble nucleation must be confined 
very close to the symmetry center. For this, we need to introduce 
some interaction between the boundary brane 
and the tunneling field $\phi$. As a simple way to introduce 
the degree of freedom that controls the strength of such an interaction, 
we consider model having a $U(\phi)$ potential localized on the brane 
as well as a potential $V(\phi)$ in the bulk. 
Define a parameter $\nu$ which controls the strength of the interaction by  
\ben
\nu:= - 4 \partial_y \log\alpha \big|_{brane} 
- 2{V'[\bar{\phi}]\over U'[\bar{\phi}]}\big|_{brane} 
+ \frac12 U''[\bar{\phi}]\big|_{brane}\,,
\een  
with $y$ being the transverse coordinate to the brane.    
Then, on the assumption that the effect of the gravitational back reaction 
is small, perturbing the most symmetric (i.e., $\Delta=0$ case) 
instanton solution $\bar{\phi}$, we can estimate the probability distribution 
of the off-centred bubble nucleation as a function of $\Delta$:    
\ben
  P(\Delta)\propto e^{-S_E[\bar{\phi}]}
          \exp \left( 
                     -\frac12 \alpha_B^4 \nu M^5 \Delta^2 
               \right) \,. 
\een
Here $S_E[\bar{\phi}]$ is the Euclidean action for the symmetric 
solution~$\bar{\phi}$, and $M$ denotes an energy scale 
of the bulk scalar field, being related to the difference of 
the vacuum energy between true and false vacua: $\delta V \approx M^5$. 
From this, we can find that the required 
concentration~(\ref{condi:flatness}) 
of the nucleation point can be realized if the tunneling field has 
an appropriate potential localized on the brane.  

However, we have to care the following point. 
The interaction between the brane and the tunneling field 
inevitably increases the effective mass-squared of the perturbation 
modes corresponding to bubble wall fluctuations.   
When there is no interaction, this effective mass squared is 
negative and the wall fluctuation grows until it hits the brane. 
If this mass squared becomes positive, 
the fluctuation modes are stabilized, and the bubble wall 
never hit the brane. Hence, the interaction should be strong enough
to force the bubble nucleate near the center of the symmetry 
but weak enough to let the bubble fluctuation grow.  
Our model thus requires one fine tuning to adjust the strength 
of this interaction. 

\begin{figure}  
  \begin{center}
    \begin{minipage}{7cm}
  \centerline{\epsfxsize = 7cm \epsfbox{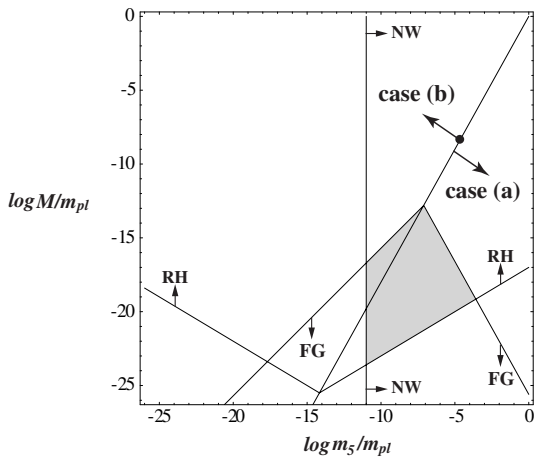}} 
    \end{minipage}
  \end{center}
\begin{center}
\begin{minipage}{14cm}
        \caption{\small   
                  The constraints on the parameters $(m_5,M)$ are
                  shown in the units of $m_{pl}$, where $m_5$ is 
                  $5$-dimensional Planck mass. 
                  The parameters in the shaded region are allowed for 
                  $O(\nu\ell)\approx 1$. 
                  {\bf FG} denotes the constraint that the bubble collide 
                  with the brane and result in the sufficiently flat 
                  universe, and {\bf NW} denotes the constraint that 
                  the Newton's' law be valid on scale larger than 1mm. 
                  {\bf RH} is the constraint on the reheating temperature.  
                 }  
\protect  \label{fig:Mconstraint} 
\end{minipage}
\end{center}  
\end{figure}

These constraints on the model parameters are summarized in 
Fig.~\ref{fig:Mconstraint}, where besides the conditions mentioned above,  
we also have taken into account the following two requirements: 
sufficiently high reheating temperature 
for the standard nucleosynthesis to proceed successfully, 
and the recovery of Newton's law up to 1mm. 

\section{Discussion} 
\label{Sect:Discussion} 

Concerning the constraints for the model parameters, 
we can find in Fig.~\ref{fig:Mconstraint} that still a wide region 
in the parameter space is not excluded. 
However, the result must interpreted carefully. 
The interaction strength needs to be tuned additionary. 
Unfortunately we could not find a natural explanation for this parameter 
tuning within our simple model. We expect future new invention on this point. 

In Ref.~\cite{GIT}, the amplitude of density fluctuations of this model 
was not estimated. The analysis of density perturbations will bring another 
meaningful constraint on the model parameters.  
To apply a similar analysis done for Bucher's model~\cite{JB,GT2} 
to this model, however further extension of the formalism is necessary. 

In the sense that our universe is realized inside a single nucleated 
bubble, our new scenario has a common feature with 
the one-bubble open inflation~\cite{oboi-1,oboi-2}. 
However, in the one-bubble open inflation, 
since the bubble wall, namely, the boundary surface of the old inflationary 
phase, is timelike, the universe inside the nucleated bubble 
is curvature dominant from the beginning. 
Hence the flatness problem is not solved 
without the second inflationary epoch. 
On the other hand, in the present scenario, 
the bubble nucleation occurs not on the brane but inside the bulk. 
The boundary surface of the inflation on the brane is spacelike, 
hence the brane universe can be sufficiently flat 
without introducing a second inflation. 

If we interpret our scenario on the viewpoint of the 4-dimensional 
effective theory, it does not look quite natural. The phase transition 
occurs beyond the horizon scale in a completely synchronized manner. 
It will be necessary to consider a slightly complicated situation 
in order to explain such a process without assuming the 
existence of extra-dimension(s). 
This means that our scenario gives a new paradigm
opened for the first time in the context of the braneworld.

As a simple case, we considered a single bubble nucleation in the bulk. 
There is, however, a possibility that many bubbles nucleate. 
If nucleation of multi bubbles can also be confined near the bulk 
symmetric center, then the situation will be similar to the single 
bubble case. As vacuum bubbles expand, they immediately collide with 
each other after the nucleation and continue to expand as a single bubble. 
The bubble collision may produce inhomogeneities on the bubble wall. 
However the rapid expansion of the bubble wall will erase 
such inhomogeneities by the time of the brane big-bang. 
If the bulk field potential $V(\phi)$ has a number of different vacua and 
bubbles are nucleated in the different vacuum phases, the collision of 
the nucleated bubbles may produce topological defects of lower dimension, 
which could generate large inhomogeneity on the brane universe through 
the collision. But provided again that the nucleation of bubbles 
is confined very near the center, one can expect that the abundance of 
such lower dimensional topological defects in a horizon scale of 
the FLRW brane is reduced 
sufficiently by the de Sitter-like expansion of the bubble wall, 
as the standard inflation solves the problem of unwanted relics. 

Our scenario has a lot in common with Bucher's model~\cite{Bucher} 
in the sense that the bubble 
collision in higher dimensional spacetime brings a big-bang 
to our 4-dimensional world realized on the brane. 
Although the basic idea is quite similar, these two models have 
many different aspects. 
In Bucher's model, 
the flatness problem is solved by the large separation 
between the nucleation centers of the two bubbles 
when the initial bulk is Minkowskian.    
When the initial bulk is de Sitter space,
the two bubble centers must be located at almost 
anti-podal points of the bulk each other. 
The place of collision is not special at all before the 
collision occurs, and hence the tension of the brane formed 
after the bubble collision is brought by the colliding bubbles.  
On the other hand, in our model the universe starts with a small 
bulk initially, and the brane with positive tension exists 
as a target for the bubble wall to collide. As a consequence of this, 
the localization of the nucleation center of the colliding bubble 
became necessary instead of the large separation of the bubbles, 
or requirement for the two bubbles to locate at almost anti-podal points 
each other.  
%

Following point is thought to be a common problem arising 
in any cosmological colliding brane models. 
The mechanism of energy transfer at brane collision process 
has not been made clear yet. 
In our model, we simply assumed that the energy of the bubble 
is completely converted into the energy of radiation on the brane.  
However, the process of collision will be heavily model dependent, 
and it is easy to imagine other possibilities. For example, 
the collision might be elastic, and then the bubble bounces 
into the bulk. The energy of the bubble wall may completely dissipate 
into the bulk as radiation. 
The energy dissipating into the bulk may produce the Weyl components 
of the bulk gravity. 
Then, they affect the evolution of the FLRW brane as dark radiation,  
whose energy density must be suppressed compared to that of 
the fields localized on the brane. 
Otherwise, the standard big-bang nucleosynthesis would not work. 
To avoid this problem, our scenario may require
a certain mechanism which realizes the efficient energy conversion 
from the bubble wall to the fields localized on the brane. 
This might be possible, 
for example, if a sufficiently large number of light fields 
which couple to the bulk scalar field reside on the brane. 
If the inverse of the bulk curvature radius is larger 
than the reheating temperature, most of the KK modes of 
the bulk fields will not be excited by the collision. 
Then, the number of relevant degrees of freedom localized on the brane 
is larger than that in the bulk. In such a situation,  
once the equi-partition among these relevant degrees of freedom 
is established, 
the relative contribution from dark radiation is suppressed. 
Alternatively, we might be able to construct a model in which 
such a relic dark radiation is diluted by 
a fairly short period inflation like thermal 
inflation~\cite{ThermalInflation} 
implemented by the potential on the brane. 
To further investigate this issue, 
one needs to specify the details of the model,    
which will be supplied once we can embed brane collision model in 
more fundamental theories such as string theory or M-theory.

\section*{\bf Acknowledgments}         

We would like to thank H. Kodama, K. Kohri and K. Nakao 
for comments and discussion. We would like to express special thanks to 
J. Garriga, who suggested several important improvements~\cite{private}. 
A.I. and U.G. are supported by Japan Society for the Promotion of Science.  
The work by T.T. is supported in part by 
Monbukagakusho Grant-in-Aid No.~1270154 and Yamada foundation. 
The authors thank the Yukawa Institute for Theoretical Physics 
at Kyoto University. Discussions during the YITP workshop YITP-W-01-15 
on ``~Braneworld - Dynamics of spacetime with boundary~'' 
were also very useful to complete this work.


\begin{thebibliography}{99}

\bibitem{GIT} 
U.~Gen, A.~Ishibashi, and T.~Tanaka, hep-th/0110286.  

\bibitem{HW} 
P.~Ho\v rava and E.~Witten, 
Nucl. Phys. {\bf B460}, 506 (1996); {\bf B475}, 94 (1996).   

\bibitem{RS}   
L.~Randall and R.~Sundrum,
Phys.\ Rev.\ Lett.\ {\bf 83}, 3370 (1999);  
L.~Randall and R.~Sundrum,  
Phys.\ Rev.\ Lett.\ {\bf 83}, 4690 (1999). 

\bibitem{BWC} 
P.~Binetruy, C.~Deffayet, and D.~Langlois, 
Nucl. Phys. B {\bf 565}, 269 (2000); 
J.M.~Cline, C.~Grojean, and G.~Servant, 
Phys. Rev. Lett. {\bf 83}, 4245 (1999); 
P.~Kraus, JHEP 9912 (1999) 011; 
P.~Binetruy, C.~Deffayet, U.~Ellwanger, and D.~Langlois, 
Phys. Lett. B {\bf 477} 285 (2000); 
A.Y.~Neronov, Phys.~Lett. B {\bf 472}, 273 (2000); 
S.~Mukohyama, Phys.~Lett. B {\bf 473}, 241 (2000); 
D.~Ida, JHEP 0009 (2000) 014; 
P.~Binetruy, C.~Deffayet, U.~Ellwanger, and D.~Langlois, 
Phys. Lett. B {\bf 477}, 285 (2000); 
E.E.~Flanagan, S.-H.~Henry Tye, and I.~Wasserman, 
Phys. Rev. D {\bf 62}, 024011 (2000); 
C.~Csaki, M.~Graesser, C.~Kolda, and J.~Terning, 
Phys. Lett. B {\bf 462}, 34 (1999); 
C.~Csaki, M.~Graesser, L.~Randall, and J.~Terning, 
Phys. Rev. D {\bf 62}, 045015 (2000); 
S.~Mukohyama, T.~Shiromizu, and K.~Maeda, 
Phys. Rev. D {\bf 62}, 024028 (2000); 
Erratum-ibid. D {\bf 63}, 029901 (2001); 
P.~Bowcock, C.~Charmousis, and R.~Gregory, 
Class. Quant. Qrav. {\bf 17}, 4745 (2000); 
L.~Anchordoqui, C.~Nunez, and K. Olsen, 
JHEP 0010, 050 (2000); 
L.~Anchordoqui, J.~Edelstein, C.~Nunez, S.P.~Bergliaffa, 
M.~Schvellinger, M.~Trobo, and F.~Zyserman, 
Phys. Rev. D {\bf 64} 084027 (2001). 

\bibitem{CP} 
H.~Kodama, A.~Ishibashi and O.~Seto, 
Phys.\ Rev.\ D {\bf 62}, 064022 (2000);  
S.~Mukohyama, 
Phys.\ Rev.\ D {\bf 62}, 084015 (2000); 
S.~Mukohyama,  
Class.\ Quant.\ Grav.\ {\bf 17}, 4777 (2000); 
R.~Maartens,
Phys.\ Rev.\ D {\bf 62}, 084023 (2000); 
K.~Koyama and J.~Soda, 
Phys.\ Rev.\ D {\bf 62}, 123502 (2000); 
C.~van de Bruck, M.~Dorca, R.~H.~Brandenberger and A.~Lukas, 
Phys.\ Rev.\ D {\bf 62} 123515 (2000);  
D.~Langlois, 
Phys.\ Rev.\ D {\bf 62}, 126012 (2000); 
N.~Deruelle, T.~Dolezel, and J.~Katz, 
Phys.\ Rev.\ D {\bf 63}, 083513 (2001); 
D.~Langlois, R.~Maartens, M.~Sasaki, and D.~Wands, 
Phys.\ Rev.\ D {\bf 63} 084009 (2001); 
H.A.~Bridgman, K.A.~Malik, and D.~Wands, 
astro-ph/0107245; 
K.~Koyama and J.~Soda, 
hep-th/0108003; 
P.~Brax, C.~van~de~Bruck, and A.C.~Davis, 
hep-th/0108215. 

\bibitem{braneinflation}
G.~Dvali, S.-H.~Henry Tye, 
Phys.\ Lett.\ B {\bf 450} 72 (1999);
C.P.~Burgess, M.~Majumdar, D.~Nolte, F.~Quevedo, G.~Rajesh, R.-J.~Zhang,
JHEP {\bf 0107} 047 (2001);
G.R.~Dvali, Q.~Shafi and S.~Solganik, hep-th/0105203;
J. Garc\'{\i}a-Bellido, R. Rabadan, F. Zamora, JHEP {\bf 0201} 036 (2002); 
G.~Shiu and S.~H.~Tye, Phys.\ Lett.\ B {\bf 516} 421 (2001); 
C. Herdeiro, S. Hirano, R. Kallosh, JHEP {\bf 0112} 027 (2001); 
B.~s.~Kyae and Q.~Shafi, ``Branes and inflationary cosmology, Phys.\
	Lett.\ B {\bf 526} 379 (2002); 
C.P. Burgess, P. Martineau, F. Quevedo, G. Rajesh, R.-J. Zhang,
JHEP {\bf 0203} 052 (2002); 
R. Blumenhagen, B. Kors, D. L\"ust, T. Ott,
hep-th/0202124;
K. Dasgupta, C. Herdeiro, S. Hirano, R. Kallosh,
hep-th/0203019;
N. Jones, H. Stoica, S.-H.Henry Tye, 
ep-th/0203163.


\bibitem{KOST}
J. Khoury, B.A. Ovrut, P.J. Steinhardt, and N. Turok, 
Phys. Rev. D {\bf 64}, 123522 (2001). 

J.~Khoury, B.A.~Ovrut, P.J.~Steinhardt, and N.~Turok, 
hep-th/0109050.  


\bibitem{KKL} 
R. Kallosh, L. Kofman, and A. Linde, 
Phys. Rev. D {\bf 64}, 123523 (2001).  

\bibitem{ControversialKOST} 
D.H.~Lyth,
Phys. Lett. B {\bf 526}, 173 (2002); 
R.H.~Brandenberger and F.~Finelli, JHEP {\bf 0111}, 056 (2001), 
F.~Finelli and R.H.~Brandenberger, hep-th/0112249; 
J.~Hwang, Phys. Rev. D {\bf 65}, 063514 (2002); 
J.~Hwang and H.~Noh, astro-ph/0221079; 
S.~Tsujikawa, 
Phys. Lett. B {\bf 526}, 176 (2002); 
J.~Martin, P.~Peter, N.~Pinto-Neto, and D.J.~Schwarz, 
hep-th/0112128; 
J.~Martin, P.~Peter, N.~Pinto-Neto, and D.J.~Schwarz, 
hep-th/0204222.  

\bibitem{BGCGLP} 
M.~Bastero-Gil, E.J.~Copeland, J.~Gray, A.~Lukas, and M.~Plumacher, 
hep-th/0201040.  

\bibitem{bubblenucleation}
W.B.~Perkins, 
Phys. Lett. B {\bf 504}, 28 (2001); 
R.~Cordero and A.~Vilenkin, 
hep-th/0107175. 

\bibitem{Gorsky}
A.~Gorsky and K.~Selivanov, 
Phys. Lett. B {\bf 485}, 271 (2000); 
A.S.~Gorsky and K.G.~Selivanov, 
hep-th/0006044. 

\bibitem{Bucher} 
M.~Bucher, 
Phys. Lett. B {\bf 530}, 1 (2002).  

\bibitem{CD} 
S.~Coleman and F.~De Luccia, Phys. Rev. D {\bf 21}, 3305 (1980). 

\bibitem{JB} 
J.J.~Blanco-Pillado and M.~Bucher, Phys. Rev. D {\bf 65}, 083517 (2002).  

\bibitem{GT2}
J.~Garriga and T.~Tanaka, Phys. Rev. D {\bf 65}, 103506 (2002).  

\bibitem{collision} 
T.~Dray and G.~{}'t Hooft, Commun. Math. Phys. {\bf 99}, 613 (1985); 
I.H.~Redmount, Prog. Theor. Phys. {\bf 73}, 1401 (1985); 
E.~Poisson and W.~Israel, Phys. Rev. D {\bf 41}, 1796 (1990); 
D.~Nunez, H.P.~de Oliveira, and J.~Salim, 
Class. Quant. Grav. {\bf 10}, 1117 (1993); 
K.~Nakao, D.~Ida, and N.~Sugiura, Prog. Theor. Phys. {\bf 101}, 47 (1999); 
D.~Ida and K.~Nakao, Prog. Theor. Phys. {\bf 101}, 989 (1999).  

\bibitem{LMW}
A.~Neronov, hep-th/0109090;   
D.~Langlois, K.~Maeda, and D.~Wands, 
Phys. Rev. Lett. {\bf 88}, 181301 (2002).  

\bibitem{CGM}
S.~Coleman, V.~Glaser, and A.~Martin, 
Commun. Math. Phys. {\bf 58}, 211 (1978). 


\bibitem{oboi-1} 
M.~Bucher, A.S.~Goldhaber, and N.~Turok, Phys. Rev. D {\bf 52}, 3314 (1995).  
\bibitem{oboi-2} 
K.~Yamamoto, M.~Sasaki and T.~Tanaka, Astrophys. J. {\bf 455}, 412 (1995).


\bibitem{ThermalInflation} 
D.H.~Lyth and E.D.~Stewart, Phys. Rev. Lett. {\bf 75}, 201 (1995); 
D.H.~Lyth and E.D.~Stewart, Phys. Rev. D {\bf 53}, 1784 (1996).  

\bibitem{private}
J.~Garriga, {\it private communication} : 
The expression for the Lorentz 
factor in the earlier version 
of the paper~\cite{GIT} is valid only for the case of Minkowski bulk. 
This mistake was pointed out. 
The importance of the condition for the existence of growing 
modes discussed in Sec.~2 was also pointed out.  

\end{thebibliography}
\end{document}